\documentclass[article,printer]{aa}
\usepackage{graphicx}
\usepackage{aas_macros}
\usepackage{natbib}
\usepackage{amssymb}
\usepackage{amsmath}
\usepackage{txfonts}
\def\src{1RXS\,J170849.0$-$400910}
\def\xmm {\emph{XMM-Newton}}
\def\cxo {\emph{Chandra}}

\def\sax {\emph{BeppoSAX}}
\def\xte {\emph{RXTE}}
\def\int {\emph{INTEGRAL}}

\def\swi {\emph{Swift}} 
\begin{document}
   \title{Long term hard X-ray variability of the anomalous X-ray pulsar 1RXS J170849.0--400910 discovered with \int}

   \author{D. G\"{o}tz\inst{1}, N. Rea\inst{2,9}, G.L. Israel\inst{3}, S. Zane\inst{4}, P. Esposito\inst{5,6}, E.V. Gotthelf\inst{7},
                S. Mereghetti\inst{6}, A. Tiengo\inst{6}, \and R. Turolla\inst{8,4}
          }

   \offprints{diego.gotz@cea.fr}

   \institute{CEA Saclay, DSM/Dapnia/Service d'Astrophysique, F-91191, Gif sur Yvette, France
\and
SRON--Netherlands Institute for Space Research, Sorbonnelaan 2, 3584 CA, Utrecht, The Netherlands
\and
INAF--Osservatorio Astronomico di Roma, Via Frascati 33, I-00040 Monteporzio Catone (Roma), Italy
\and
Mullard Space Science Laboratory, University College London, Holmbury St. Mary, Droking Surrey, RH5 6NT, UK
\and
Universit\`a degli Studi di Pavia, Dipartimento di Fisica Nucleare e Teorica and INFN-Pavia, via Bassi 6, I-27100 Pavia, Italy
\and
INAF--Istituto di Astrofisica Spaziale e Fisica Cosmica Milano, via Bassini 15, I-20133 Milano, Italy
\and
Columbia Astrophysics Laboratory, Columbia University, 550 West 120th Street, New York, NY 10027, USA
\and
University of Padua, Department of Physics, via Marzolo 8, 35131 Padova, Italy
\and
University of Amsterdam, Astronomical Institute "Anton Pannekoek", Kruislaan 403, 1098 SJ, Amsterdam, The Netherlands 
}

   \date{Received 17/07/2007 / accepted 17/09/2007}
\authorrunning{G\"{o}tz et al.}
\titlerunning{\src}

  \abstract
{}
{ We report on a multi-band high-energy observing campaign
aimed at studying the long term spectral variability of the
Anomalous X-ray Pulsar (AXP) \src, one of the magnetar
candidates.}
   { We observed \src\ in Fall 2006 and Spring 2007 simultaneously with \swi/XRT,
   in the \mbox{0.1--10 keV} energy range, and with \int/IBIS, in the 20--200 keV energy range. Furthermore,
   we also reanalyzed, using the latest calibration and software, all the publicly available \int\ data since
   2002, and the soft X-ray data starting from 1999 taken using \sax, \cxo,
   \xmm\, and  \swi/XRT, in order to study the soft and hard X-ray spectral variability of \src.}
   {We find a long-term variability of the hard X-ray flux, extending the hardness-intensity
   correlation proposed for this source over 2 orders of magnitude in energy.}
   {}


  \maketitle
%

\section{Introduction}
Anomalous X-ray Pulsars \citep[AXPs,][]{mereghetti02_axp} are a
peculiar subclass of X-ray pulsars which share the following
properties: a narrow range of spin periods ($P=5$--12 s), a typical
X-ray luminosity of $L_{\rm{X}}\sim 10^{34}$--10$^{36}$ erg s$^{-1}$, no
evidence for Doppler shifts in the light curve, the lack of bright
optical companions -- indicating their isolated nature -- and a
spin down in the range 10$^{-13}$--10$^{-10}$ s s$^{-1}$. 
Their soft X-ray spectra are generally well described by a two
component model, made of a black body with $kT\sim$ 0.4--0.5\,keV,
and a steep power law
\citep[see][and references therein]{woods06}. The
detection of Soft Gamma-ray Repeater-like bursts from five AXPs
\citep[e.g.][]{gavriil02} has pushed forward the
interpretation of the AXPs as magnetars, i.e. neutron stars with
an ultra-high magnetic field (B $\sim$10$^{14-15}$ G). The
magnetar model \citep{duncan92}, developed initially to describe the
SGRs phenomenology (bursts, timing, persistent emission), seems
quite capable to describe the AXPs characteristics as well.\\
\indent Since the AXPs spectra below $\sim$10 keV are rather soft, the
first \int\ detections above 20 keV of very hard high-energy tails
associated with these objects came as a surprise
\citep{kuiper06,denhartog06,revnivtsev04,mgm05,molkov05,gotz06}. 
AXP spectra flatten ($\Gamma\sim$ 1)
above 20 keV and the pulsed fraction of some of them reaches up to 100\% \citep{kuiper06}. 
The discovery of these hard tails provides new constraints on the emission models for these objects
since their luminosities might well be dominated by hard, rather than soft, X-rays.\\
\indent \src\ was discovered during the {\it ROSAT} all sky survey. The measure of
its period \citep{sugizaki97}, period derivative \citep{kg03} and
general X-ray properties \citep{ics99}, made it an AXP member.
Interesting results have been reported by \cite{roz05}, who claimed a
correlation between the soft X-ray flux and spectral hardness, with a marginal evidence of
the highest and hardest spectra being correlated with the AXP glitching activity
\citep{dallosso03,kaspi00,israel07}.
This correlation has been recently confirmed by \cite{campana07}, \cite{rio07},
and \cite{israel07} using further \cxo, \swi/XRT and \xte\ data,
but always focussing on a limited energy interval (i.e. below $\sim$10 keV).\\
\indent In this paper, we present our recent multi-wavelength observation
campaign carried out with \int\ \citep{winkler03} and \swi\
\citep{gehrels04}. These new data, together with a re-analysis of all
publicly available \int, and soft X-ray observations of \src,
allow us to investigate the timing and spectral properties of the
source over a broader energy range. For our timing analysis we made use of the solution
recently derived by \citet{israel07}, see Table \ref{tab:time}.
We report a long-term correlation between
the soft and hard X-ray emission. Results are discussed in the framework of the magnetar model.

\section{Observations and Data Reduction}

\subsection{\int /IBIS observations}

We selected and analyzed all publicly available IBIS \citep[\int\ coded
mask imager;][]{ubertini03} pointings within $12\degr$ from the direction of
the source, for a total of 2550 pointings of 2--3 ks each. In addition, we
analyzed our data of the Key Programme observation of the Galactic Centre,
622 pointings performed in Fall 2006, and Spring 2007, since, thanks to
its large field of view (29$\degr\times$29$\degr$), the source was
almost always covered during these observations.  Our analysis is based on
data taken with ISGRI \citep{lebrun03}, the IBIS low energy detector array,
which is made of CdTe crystals, and is working in the 15 keV--1 MeV
energy range.\\
\indent We processed the data using the Offline Scientific Analysis (OSA) software
provided by the \int\ Science Data Centre (ISDC, \citealt{courvoisier03}) v6.0. We
produced the images of each pointing in ten energy bands between 20 and \mbox{300
keV}. We then added up the images in order to detect the source in the
20--70 keV energy range.
To investigate the broad-band temporal X-ray variability, the observations
were ganged-up to overlap available soft X-ray data sets.
The six data segments used in this study are shown in Table \ref{tab:int}.
From each of these segments we derived spectra from the mosaicked
images in the 10 energy bands.
\begin{table}[ht!]
\begin{minipage}{\columnwidth}
\caption{\int\ observations and results. Errors are at 1\,$\sigma$ c.l.} 
\label{tab:int}
\centering
\begin{tabular}{cccccc}
\hline  \hline
Obs. & T$_{Start}$ & T$_{Stop}$ & Exposure & Type$^{\mathrm{a}}$ &20--70 keV\\ 
 Number & MJD & MJD & Ms &  & Flux$^{\mathrm{b}}$\\ 
\hline
1 & 52698.32 & 52751.59 & 0.83 & P &0.41$\pm$0.06\\ 
2 &  52859.63 & 52919.88 & 1.16 &  P&0.27$\pm$0.06\\
3 &  53226.18 & 53298.48  & 1.0  & P &0.42$\pm$0.06\\ 
4 &  53408.32& 53481.27 & 1.5 & P&0.44$\pm$0.05 \\ 
5 & 53990.59 & 54013.77 & 0.46 & KP&0.16$\pm$0.09 \\  
6 & 54159.36 & 54184.07 & 0.52 & KP&0.23$\pm$0.09 \\ 
\hline 
\end{tabular}
\end{minipage}
\begin{list}{}{}
\item[$^{\mathrm{a}}$] P = Public ; KP = Key Programme.
\item[$^{\mathrm{b}}$] In units of counts s$^{-1}$ on the ISGRI camera.
\end{list}
\end{table}
\vspace{-1cm}

\subsection{X-ray observations}

We re-analysed all the public available recent data of \src\ taken with
X-ray imaging telescopes in the soft X-rays (i.e. below $\sim$10 keV).
The observation times and exposures for each
data set are reported in Table \ref{tab:xray}.\\
\indent \sax, \xmm\ (PN), and \cxo\
data were reduced following the procedures described in
\citet{israel01,rea03,roz05,campana07},  but using the latest available software and
calibration files.\\
\indent For the \swi/XRT we analysed only data taken in photon counting
(PC) mode. These were reduced using the FTOOLS v6.2 package and
CALDB calibration files v20070531 provided by the High Energy Astrophysics
Science Archive Research Center.  Standard cleaning and selections were applied and
spectra were extracted using regions of 40$^{\prime\prime}$ and 80$^{\prime\prime}$ radius
for source and background, respectively. Standard response files have been
used (we used v.9 response matrices).\\
\indent We notice that \sax\ observed the source also in the 15--300 keV
band, with
the PDS instrument
\citep{frontera97}. This non-imaging spectrometer had a field of view
of $1.3\degr$ (FWHM) and the background subtraction was done with a
rocking system, which switched between the source and two offset
background regions. The presence of bright contaminating sources in both the background
pointings prevented us from analyzing the background subtracted spectra,
so we concentrated on the timing analysis of these data and
extracted barycentered events files for the on-source position (see
Sect.\,\ref{timing}).
\begin{table}[ht!]
\begin{minipage}{\columnwidth}
\caption{X-ray data observations and results. Errors are at 1\,$\sigma$ c.l.} 
\label{tab:xray}
\centering
\begin{tabular}{ccccc}
\hline  \hline
Instrument & OBS date & Exposure & 1--10 KeV                             & Photon\\ 
Name         & MJD         & ks            & Flux$^{\mathrm{a}}$ &Index\\ 
\hline  
\emph{SAX} {\footnotesize LECS/MECS} & 51268 & 26/52 &
4.40$^{+0.03}_{-0.03}$ &2.63$^{+0.07}_{-0.07}$\\
\emph{SAX} {\footnotesize LECS/MECS} & 52138 & 77/199 &
4.45$^{+0.01}_{-0.02}$ &2.38$^{+0.03}_{-0.04}$\\
\cxo\ {\footnotesize HETG} &  52526 & 32 &
3.9$^{+0.1}_{-0.2}$ &2.39$^{+0.12}_{-0.12}$\\
\xmm\ {\footnotesize PN} &  52879 & 31 &
3.00$^{+0.01}_{-0.01}$ & 2.77$^{+0.01}_{-0.01}$  \\
\cxo\ {\footnotesize CC} &  53189 &  29 &
3.82$^{+0.08}_{-0.10}$ &2.74$^{+0.04}_{-0.08}$ \\
\swi\ {\footnotesize XRT} & 53425 & 12 &
4.28$^{+0.11}_{-0.08}$ &2.47$^{+0.06}_{-0.06}$\\
\swi\ {\footnotesize XRT} & 54017  &  9 &
2.88$^{+0.09}_{-0.09}$ &2.99$^{+0.05}_{-0.05}$ \\
\swi\ {\footnotesize XRT} & 54182 &  4 &
3.26$^{+0.12}_{-0.13}$ &2.76$^{+0.08}_{-0.06}$\\
\hline
\end{tabular}
\end{minipage}
\begin{list}{}{}
\item[$^{\mathrm{a}}$]Absorbed flux in units of 10$^{-11}$erg cm$^{-2}$ s$^{-1}$.
\end{list}
\end{table}
\vspace{-1.cm}

\begin{figure*}[t]
\centering
\hbox{
\hspace{1.cm}
\includegraphics[width=7cm]{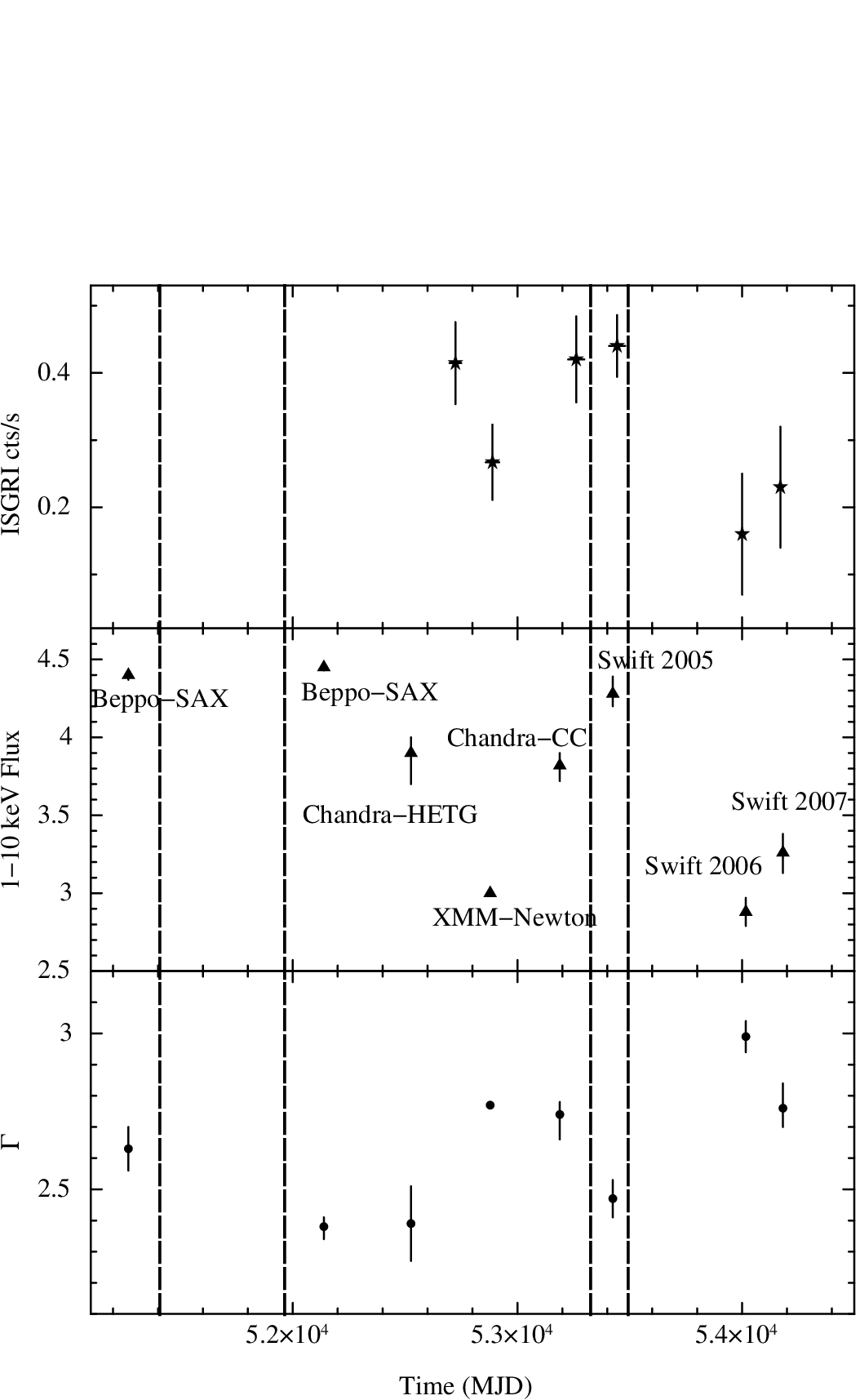}
\hspace{1.5cm}
\includegraphics[width=7cm]{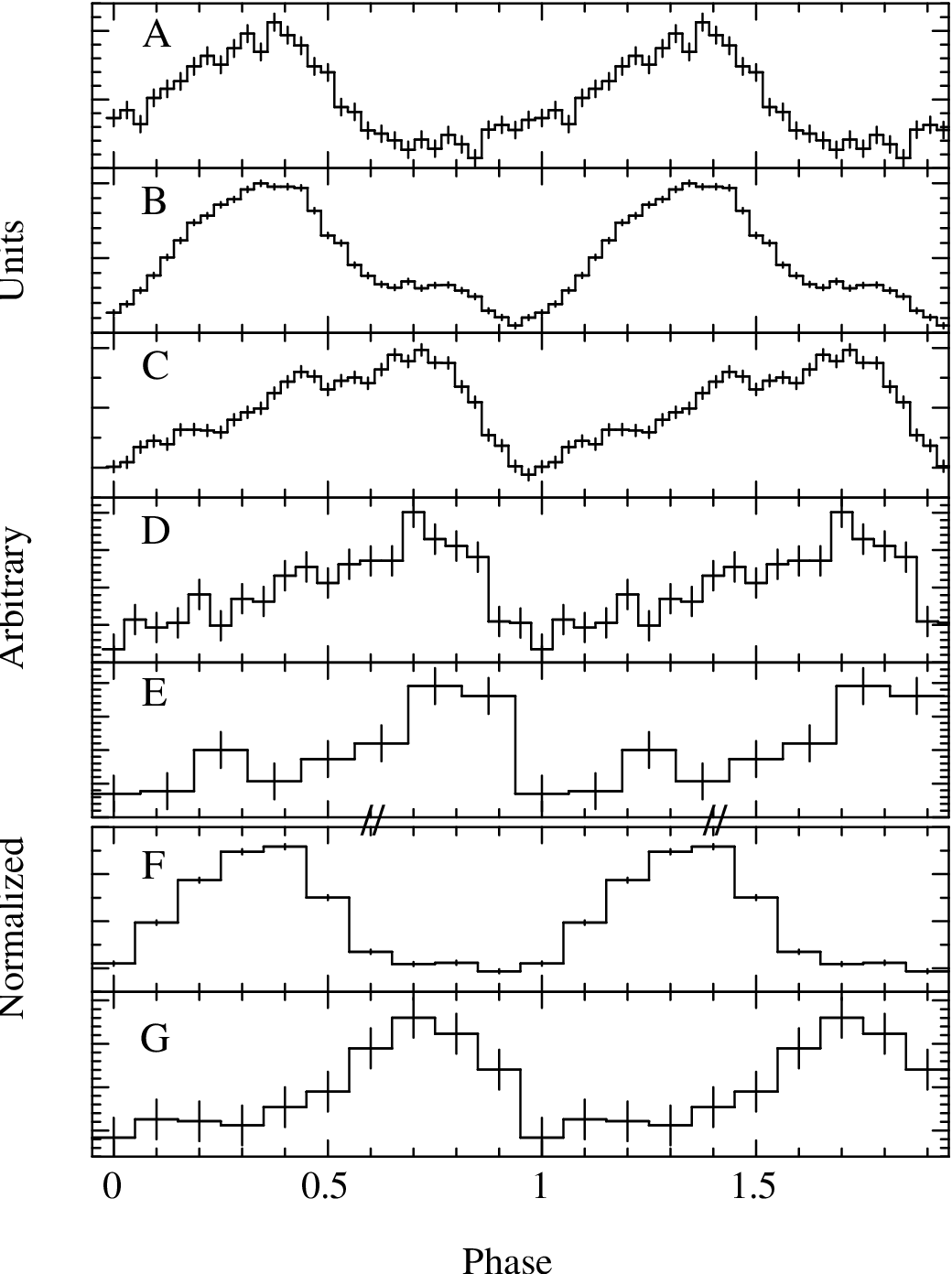}  }
\caption{{\em Left plot}. Upper Panel: hard X-ray fluxes derived with \int/IBIS (20--70 keV). Middle Panel: absorbed 1--10 keV fluxes in units of 10$^{-11}$ erg cm$^{-2}$ s$^{-1}$ derived from recent observations of
X-ray imaging telescopes as a function of time. Lower Panel: photon indices measured in the 1--10 keV energy band.
Vertical dashed lines mark the times of four observed glitches,
\citep[see][and references therein]{israel07}. {\em Right plot}. Folded light curves of \src, obtained with \xte/PCA (panels
A--D; using the whole 2004 available data) and IBIS data (obs.\,3) folded on the \xte\ timing solution (panel E, data from
obs.\,3).
The energy bands are, from top to bottom: 2.5--4, 4--8, 8--16, 16--32, 20--200
keV. Panels F and G represent the MECS (1--10 keV) and PDS (20--200 keV)
light curves registered during the second (2001) \sax\
observation. Note
that panels F and G are phase-aligned between themselves, but not with the
others. Errors are for both plots reported at 1\,$\sigma$ c.l.}
\label{fig:corr}
\end{figure*}

\section{Analysis and Results}
\label{sec:results}
\subsection{Spectral analysis}
\label{spectra}
\begin{figure}[ht]
\centering
\includegraphics[width=6cm,angle=-90]{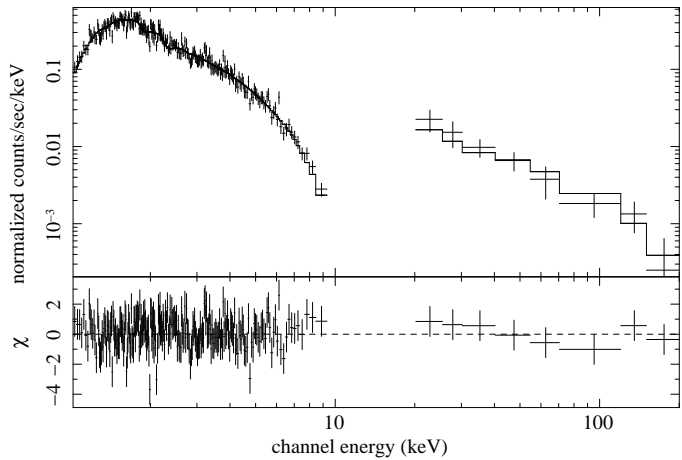}
\caption{2005 ``high" state phase averaged spectrum of \src. Upper panel: the points below 10 keV are taken with \swi/XRT, while above 20 keV IBIS/ISGRI data points are shown. The continuous line represents the best fit model (RCS + power law) described in the text. Lower panel: residuals with respect to the best fit model, plotted as a dashed line.}
\label{fig:spectrum}
\end{figure}

The spectral parameters reported in Table \ref{tab:xray} have been derived
by fitting all the datasets simultaneously in 1--10 keV energy range (except for
{\em Chandra} data which were limited to 8\,keV) by using an absorbed black body plus power
law model, with a multiplicative constant factor taking into account inter-calibration issues,
not allowed to vary for the same instrument and observing mode.
While the parameters of the power law have been left free to vary, the
absorption column density, and the black body temperature have been forced to be the same
for all instruments. We found a good fit ($\chi^2$/d.o.f. =  1188/1264 = 0.94),
and the derived values were $N_{\rm{H}}=1.36(1)\times10^{22}$ cm$^{-2}$ 
(solar abundances assumed from \citealt{anders89}, and photoelectric cross-section from \citealt{balucinska92}), and $kT= 0.44(1)$ keV; we verified that, if leaving these parameters free to vary, they do not
change significantly among all the observations. 
The fluxes and photon indices are reported in Table \ref{tab:xray} and Fig.\,\ref{fig:corr}\footnote{Note that all the spectral values we report here using the
latest calibrations and software are consistent with previous findings,
except for the XMM flux which, using SAS v.~7.1.0, appears slightly higher than using SAS
5.4.1 as in Rea et al.~(2005); the spectral parameters are instead
consistent.}.
Our new XRT observation campaign (last two points in Fig.\,\ref{fig:corr})
shows that the source entered a new low/soft state (similar to the one measured with {\it
XMM-Newton} in 2003), with a flux a factor $\sim$1.5 lower than
the one measured with XRT in 2005.
By adding these new points to the long
term variability study of the source, we confirm the flux-hardness
correlation proposed by \cite{roz05}. We also notice that the last three
points reported in Fig.\,\ref{fig:corr} are particularly compelling since,
being taken with the same instrument (\swi/XRT), they are not affected by
cross-calibration uncertainties.
In addition, our new hard-X data show that the long term variation in flux
is correlated over more than two orders of magnitude in energy. In
fact, IBIS observations taken quasi-simultaneously with
the last two XRT ones (i.e. in late 2006 and early 2007) indicate that
the source is hardly detected above 20 keV, while the
hard X-ray count rates measured before
followed well the variations measured in the soft X-ray range (see
Table \ref{tab:int} and Fig.\,\ref{fig:corr}).
Unfortunately, due
to the faintness of the source we could not statistically prove
spectral changes at high energies, by comparing different
\int\ observations.
In order to obtain a statistically significant high energy spectrum, we
co-added the IBIS data from observations 3 and 4. The resulting
20--200 keV spectrum is well fitted by a single power law,
without the need for a cutoff, with photon index
$\Gamma=1.46\pm0.21$. The 20--100 keV flux is
$(3.6\pm0.5)\times10^{-11}$ ergs cm$^{-2}$ s$^{-1}$.\\
\indent In the attempt to characterize the source spectrum in the low and high
state of
the source, we fitted simultaneously the IBIS observation 2 with the {\it
XMM-Newton}/PN data and the IBIS observation 4 with the XRT data. We found
that the high energy component is well above the extrapolation of the
power law derived from soft X-ray data (below $\sim$10 keV), while a three
model component, with an additional power-law describing the high energy
data points, fits well the entire broadband spectrum (1--200 keV).
Motivated by this, we tested if the spectral variations at low energies
could be induced just by the variation of the high energy power law.
Indeed, by fixing the spectral parameters derived from the {\it
XMM-Newton} observations, one can fit the broad band {\it XMM/INTEGRAL}
(2003) and {\it Swift/INTEGRAL} observations (2005) by simply changing the
slope and normalisation of the high energy power law. Unfortunately, due
to
the uncertainties in the inter-calibration, and to the low statistic of the
\swi\ data, we cannot draw a firm conclusion on this point.\\
\indent We also modeled the multi-band
spectrum in the ``low'' (2003) and ``high'' (2005) state with a
resonant cyclotron scattering model plus a power-law, and we find
a good fit in the ``low'' state with $\beta=0.49(1)$, $kT=0.36$(3),
$\tau=1.2(1)$ and $\Gamma=1.0(1)$ (see Rea et al. 2007b for
details on the model). In the ``high'' state, $\beta$ and $kT$ do
not vary, while we found a $\tau=1.8(1)$ and $\Gamma=0.8(1)$,
which reflects the hardening of the spectrum. The ``high" state best fit
spectrum is shown in Fig. \ref{fig:spectrum}.

\subsection{Timing analysis}
\label{timing}
IBIS data alone do not allow a blind search
for timing signatures. In order to perform a timing analysis in the hard
X-ray range, we first derived a phase coherent timing solution using
publicly available \xte/PCA data covering the same period of the IBIS data.
We discovered two new glitches, close to the epoch of our IBIS observations 3 and 4,
which are described in detail in \citet{israel07}.
We used the temporal solutions derived there (see Table \ref{tab:time}), for the time
periods covering the \int\ data before and after the first new glitch (and
prior the second one). 
We then extracted the IBIS events in the 20--200 keV band
from the pixels that were illuminated by the source for at least the 60\%
of their surface. To optimize the signal to noise ratio, we restricted our
search to the
high state periods, namely fall 2004 and spring 2005 (obs.\,3 and 4). For obs.\,3 (obs.\,4 results are not reported, but are consistent) 
we folded the data and, by using the $Z^{2}$ test we find
that the signal is pulsed at a $\sim$5.7\,$\sigma$ level (chance
probability of $1.6\times10^{-8}$). If we divide the data in two energy
bands, 20--60 and \mbox{60--200 keV}, the significance
lowers to 3.2 and 4.2\,$\sigma$, respectively. The pulse profile of the source,
as
a function of energy, is shown in Fig.\,\ref{fig:corr}. As it can be seen,
there is a clear energy dependence of the pulse shape morphology: the peak
which is predominant in the soft band disappears and a secondary peak
grows with energy becoming the most prominent one above $\sim$8 keV. The
pulsed fraction cannot be easily derived from the PCA (a non-imaging
instrument) data due to the
uncertainties in the background estimation. Besides,
the IBIS data are not easy to handle, because of large background
variations due the to fact that
the source is at different off-axis angles during the different pointings.
In addition, due to the large IBIS field of view, the events can be
polluted by
nearby variable sources. Nevertheless, by averaging the background over
the entire IBIS observations, we could measure a pulsed fraction
by dividing the difference between the maximum and the minimum of the folded
light curves by the count rates derived from imaging. The resulting values
are: $\sim$25\% (in the 20--60 keV band), $\sim$60\% (\mbox{60--200
keV}), and $\sim$40\% (\mbox{20--200 keV}).\\
\indent We did not find significant differences in the IBIS data as far as
it concerns the pulsed profiles or the pulsed fraction between
observations 3 and 4,
i.e. before and after the third glitch reported in Fig.\,\ref{fig:corr}\\

\begin{table}[ht!]
\caption{Timing solutions used for obs.\,3 and 4, derived in \citet{israel07}}
\label{tab:time}
\centering
\begin{tabular}{cccc}
\hline  \hline
Obs. & Epoch & Period & Period derivative \\
 Number & MJD & s & 10$^{-11}$ s s$^{-1}$\\
\hline
3 & 53189.0  &11.0023066(2) & 1.915(5) \\
\hline
4 &  53189.0 &  11.002279(2) & 2.01(1)  \\
\hline
\end{tabular}
\end{table}

\indent We also reanalyzed the archival PDS data taken with \sax\ during the
pointings listed in Table\,\ref{tab:xray}.
Using the timing solutions derived from the simultaneous
MECS data \citep{rea03,israel01}
we find, and report for the first time, pulsations in the PDS data,
in the 20--200 keV energy band. Using $Z^{2}$ statistics we did not detect the pulsations 
in the first dataset, the shorter one, 
with high confidence (2.7\,$\sigma$), while in the second dataset
they were detected at 5.9\,$\sigma$.
By computing the counts in excess of the phase minimum,
we found a 20--200 keV pulsed flux of $0.025
\pm  0.010$ (for the
second observation; the results for the first one are compatible but less
significant). By assuming a power-law model with
photon index $\Gamma=1$ for the pulsed spectrum \citep{kuiper06},
this gives a flux of $\sim$ $0.7\times 10^{-11}$ erg
cm$^{-2}$ s$^{-1}$, slightly
lower but still compatible with that derived from  IBIS data by using
the same approach,
namely $\sim$ $10^{-11}$ erg cm$^{-2}$ s$^{-1}$.

\section{Discussion and Conclusions}
We have analyzed all the currently available high energy data
(\mbox{1--200 keV}) of \src\ and continued the long term variability study
of the source in the soft X-ray range ($\la$10 keV), confirming
the flux-hardness correlation proposed by \cite{roz05} and
\cite{campana07}. Moreover, we report for the first time the
discovery of hard X-ray long term flux changes, and show that
there is a possible correlation with the flux variations detected
at lower energies. Thanks to the dense monitoring of the last
years, we are able to correlate these variations with the presence
of the two new glitches \citep[discussed in][]{israel07}. The only
other magnetar for which long term variability in this band has
been reported is SGR 1806--20, for which the spectral hardening
appears to be correlated with a burst rate increase \citep{gotz07}.\\
\indent As proposed in \cite{tlk02}, \cite{roz05}, and \cite{campana07}, an
increase in the source activity (bursts, glitches) and a
simultaneous spectral hardening may be caused by a growing twist
in the magnetosphere;  this may be also responsible for
the transient appearance of a cyclotron line during the ``high''
emission state \citep{rea03,roz05}. If
this is the case, data presented here support models
in which also the hard X-ray tails are
produced by mechanisms whose strength increases with the twist. 
Indeed, it may also be that the flux variations in the hard
X-rays dominate and drive those detected at lower energies, if the
hard X-ray component does not sharply cut off below $\sim$10 keV
(as it may be the case for a synchrotron component). Quite
recently, \cite{thompson05} discussed how soft gamma-rays may be
produced in a twisted magnetosphere, proposing two different
scenarios: either thermal bremsstrahlung emission from the surface
region heated by returning currents, or synchrotron emission from
pairs created higher up ($\sim$100 km) in the magnetosphere.
While both scenarios predict a power-law-like spectral
distribution for the 20-100 keV photons, the cut-offs of the high
energy emission are markedly different in the two cases, \mbox{100 keV}
vs. 1 MeV. A third scenario involving resonant magnetic
Compton up-scattering of soft X-ray photons by a non-thermal
population of highly relativistic electrons has been proposed by
\cite{baring07}. In this contest we notice that, as for 4U 0142+614 \citep{rea07}, another member of the
AXP class, multi-band spectral observations seems to disfavor the
thermal bremsstrahlung model. In fact,  for any choice of temperature,
a model which fits well the data below $\sim$150\,keV, will over-predict by more than a factor
of 5 the COMPTEL upper limits derived in the MeV band \citep{kuiper06}. However,
these observations are not simultaneous and due to the variability of the source, no firm
conclusion can be drawn.\\
\indent IBIS data confirm the
large pulsed fraction 25\%--60\%, growing with energy, which has
been reported earlier at high energies for other AXPs (and for
\src\ using a different data set)  by \citet{kuiper06}.
There is a clear indication for a phase difference of
$\Delta\phi \sim 0.3-0.4$ between the maxima of the soft
($<$$8$ keV) and hard ($>$$8$ keV) X-rays light curves measured in the
2004 with ISGRI/PCA data (see Fig.\,\ref{fig:corr}). The same phase
shift was already present in the 2001 MECS/PDS data: within the
errors, we find no evidence for any significant evolution in
$\Delta\phi$ or pulsed flux, despite the three glitches that occurred
between the observations. The phase shift, similar to the one detected at lower energies \citep{rea03},
still lacks a clear interpretation in the magnetar framework, and it may indicate that the location of
production of the two (soft/hard) component is different, but stable with time. 

\begin{acknowledgements}

Based on observations with INTEGRAL, an ESA project.
with instruments and
science data centre funded by ESA member states (especially the PI
countries: Denmark, France, Germany, Italy, Switzerland, Spain), Czech
Republic and Poland, and with the participation of Russia and the USA.
ISGRI has been realized and maintained in flight by CEA-Saclay/DAPNIA with
the support of CNES. Based on observations with the NASA/UK/ASI Swift
mission, obtained through the High Energy Astrophysics Science Archive
Research Center Online Service, provided by the NASA/Goddard Space Flight
Center. We thank the Swift team for making these observations possible. DG
acknowledges the French Space Agency (CNES) for financial support. SZ
acknowledges STFC (ex-PPARC) for support through an AF. NR is supported by an NWO Post--doctoral 
Fellowship.

\end{acknowledgements}

\bibliographystyle{aa}
\bibliography{biblio}

\end{document}